\documentclass{IEEEtran}
\usepackage{cite}
\usepackage{amsmath,amssymb,amsfonts}
\usepackage{algorithm}
\usepackage{algorithmic}
\usepackage{graphicx}
\usepackage{textcomp}
\usepackage{tabularx,booktabs}
\newcolumntype{C}{>{\centering\arraybackslash}X} 
\def\BibTeX{{\rm B\kern-.05em{\sc i\kern-.025em b}\kern-.08em
    T\kern-.1667em\lower.7ex\hbox{E}\kern-.125emX}}
\begin{document}
\title{{A Public Key Cryptosystem Using Cyclotomic Matrices}}
\author{Md. Helal Ahmed, Jagmohan Tanti, Sumant Pushp 
\thanks{The authors are thankful and acknowledge the Council of Scientific and Industrial Research (CSIR), Government of India, for providing financial assistance to Md. Helal Ahmed.}
\thanks{Md. Helal Ahmed is with Department of Mathematics, Central University of Jharkhand, India. (e-mail: ahmed.helal@cuj.ac.in). }
\thanks{Jagmohan Tanti, is
with Department of Mathematics, Central University of Jharkhand, India. (e-mail: jagmohan.t@gmail.com).}
\thanks{Sumant Pushp is with the Department of Computer Science and Technology, Tezpur University, Assam, India (e-mail: sumantpushp@gmail.com).}}

\maketitle

\begin{abstract}
Confidentiality and Integrity are two paramount objectives in the evaluation of information and communication technology. In this paper, we propose an arithmetic approach for designing asymmetric key cryptography. Our method is based on the formulation of cyclotomic matrices correspond to the diophantine system. The proposed cyclotomic asymmetric cryptosystem (CAC) utilizes the cyclotomic matrices, whose entries are cyclotomic numbers of order $2l^{2}$, $l$ be prime over a finite field $\mathbb{F}_{p}$ of $p$ elements. The method utilize cyclotomic matrices to design a one-way function. The outcome of a one-way function that is efficient to compute however difficult to compute its inverse unless if secret data about the trapdoor is known. We demonstrate that the encryption and decryption can be efficiently performed with asymptotic complexity of $\mathcal{O}(e^{2.373})$. Besides, we study the computational complexity of the CAC.

\end{abstract}
\begin{IEEEkeywords}
Finite fields, Discrete logarithm problem, Cyclotomic numbers, Cyclotomic matrix, Public key, Secret key. 

\end{IEEEkeywords}

\section{Introduction}
\label{sec:introduction}
Apart from a rich history of Message encryption, the cryptosystem became more popular in the $20^{th}$ century upon the evolution of information technology. Until the late 1970s, all cryptographic message was transmitted by the symmetric key. This implies somebody who has enough data to encode messages likewise has enough data to decode messages. Consequently, the clients of the framework must have to impart the secret key furtively. As a result of an issue stealthily key sharing, Diffie and Hellman \cite{Hellman1} developed a totally new sort of cryptosystem called public key cryptosystem. \par
In a Public key cryptosystem, both parties (in a two-party system) have a pair of public enciphering and secret deciphering keys \cite{Stinson1, Menezes1}. Any party can send encrypted messages to a designated party using a public enciphering key. However, only the designated party can decrypt the message using their corresponding secret deciphering key \cite{Ahmad1}. After that numerous public key cryptosystems 
were presented based on tricky mathematical problems. Among these, RSA is the longest viable utilization of cryptography. In spite of the fact that since its design, despite everything it has not been broken at this point. The security of the RSA is accepted to be founded on the issue of the factorization of an enormous composite number. Be that as it may, there are some practical issues in RSA execution. The main issue is the key arrangement time that is unreasonably long for computationally restricted processors utilized in certain applications. Another issue is the size of the key. It was demonstrated \cite{Neal1} that the time required to factor an n-bit integer by \textit{index calculus factorization} technique is of order $2^{n^{1/2+\delta}}, \delta> 0$. In 1990's, J. Pollard \cite{96} demonstrated that it was possible in time bounded by $2^{n^{1/3+\delta}}, \delta> 0$. The reduction of the exponent of $n$ has significant outcomes over the long run. It should likewise be expanded each year as a result of upgrades in the factorization calculations and computational power. Until 2015, it was prescribed the base size of the RSA key should be 1024 bits and subsequently increases to 4096 \& 8192 bits by 2015 \& 2025 respectively \cite{Schneier1}. While trying to remedy these issues, Discrete logarithm problem (DLP) has been utilized (to reduce key setup time and size of the key). 
\par
Discrete logarithm problem (DLP) is a mathematical problem that occurs in many settings and it is hard to compute exponent in a known multiplicative group \cite{Meier1}. Diffie-Hellman \cite{Hellman1}
, ElGamal 
\cite{Gamal1}
, Digital Signature Algorithm \cite{SchneierBruce}, Elliptic curve cryptosystems \cite{KoblitzN, MillerV} are the schemes developed under the Discrete logarithm algorithm. The security of Diffie-Hellman relied upon the complexity of solving the discrete logarithm problem. However, this scheme has some disadvantages. It has not been demonstrated that breaking the Diffie-Hellman key exchange has relied upon DLP and also the scheme is vulnerable to a man-in-the-middle attack. For the security perspective, cryptosystem \cite{Gamal1} was proposed to introduce a digital signature algorithm (DSA) which is based on Diffie-Hellman DLP and key distribution scheme. It was demonstrated that DSA is around multiple times littler than the RSA signature and later DSA has been supplanted by the elliptic curves digital signature algorithm (ECDSA). Nonetheless, it has some practical implementation problems \cite{ECDSA, JohnsonMenezesVanstone, WienerMJ, ZuccheratoR}. The length of the smallest signature is of 320 bits, which is still being too long for computationally restricted processors. Another issue emerged is as a correlation with RSA in a field with prime characteristics, which is forty times slower than RSA \cite{EDWin}.  
\par
There are some other designs for public-key cryptosystems based on some extensive features of matrices. However, there were some practical implementation problems. Thus it had never achieved wide popularity in the cryptographic community. McElice \cite{McEliece1} proposed a public key cryptosystem based on the Goppa codes Hamming metric. The scheme has the advantage that it has two to three orders of magnitude faster than RSA. Despite its advantage, it has some drawbacks. It has demonstrated that the length of the public key is $2^{19}$ bits and the data expansion is too large. Some other extensions of the scheme can also be found in \cite{424, 1227, 976}.  Unfortunately, this scheme and its variants have been broken in \cite{1447, 1448, 882}.
Gabidulin \cite{Gabidulin1} introduced the rank metric and the Gabidulin codes over a finite field with $q$ element, where $q = p^r$ i.e. $\mathbb{F}_q$, as an alternative for the Hamming metric. The efficiency of the scheme relied upon the fact that for the same set of parameters, the complexity of the decoding algorithm for random codes in rank metric is much higher than the Hamming metric \cite{McEliece1, 5, 6, 27}. Numerous fruitful attacks were utilized on the structure of the public code \cite{17, 18, 28}. To prevent these attacks, numerous alterations of the cryptosystems were made, consequently drastically increases the size of the key \cite{12, 26, 21}. Lau and Tan \cite{Lau1} proposed new encryption with a public key matrix by considering
the addition of a random distortion matrix over $\mathbb{F}_q$ of full column rank $n$. There are also many other designs on matrices, which are not cited here, but none of them gain wide popularity in the cryptographic community due to lack of efficient implementation problems in one and another way.
 
\par
Thinking about these inadequacies, it would be desirable to have a cryptosystem dependent on other than the presumptions as of now being used. Thus we propose a cyclotomy asymmetric cryptosystem (CAC) based on strong assumptions of DLP that have to reduce the key size and faster the computational process. 

\subsection{Outline of our Scheme}
In this paper, we consider two significant problems in the theory of cyclotomic numbers over $\mathbb{F}_{q}$. The first one deals with an efficient algorithm for fast computation of all the cyclotomic numbers of order $2l^{2}$, where $l$ is prime. The subsequent one deals with designing practical public key cryptosystem based on cyclotomic matrices of order $2l^{2}$. The strategy employs for designing public-key cryptosystem utilizing cyclotomic matrices of order $2l^2$, whose entries are cyclotomic numbers of order $2l^2$, $l$ be prime, where cyclotomic numbers are certain pairs of solutions $(a,b)_{2l^2}$ of order $2l^2$ over a finite field $\mathbb{F}_{q}$ with characteristic $p$. 
\par
Cyclotomic numbers are one of the most important objects in number theory. These numbers have been extensively used in cryptography, coding theory and other branches of information theory. Thus determination of cyclotomic numbers, so called cyclotomic number problems, of different orders is one of the basic problems in number theory. Complete solutions for cyclotomic number problem for $e$ =  $2-6$, $7$, $8$, $9$, $10$, $11$, $12$, $14$, $15$, $16$, $18$, $20$, $22$, $l$, $2l$, $l^{2}$, $2l^{2}$ with $l$ an odd prime have been investigated by many authors see (\cite{Acharya1,Helal1,Helal3,Katre3,Shirolkar1} and the references there in). In our approach to designing cyclotomy asymmetric cryptosystem (CAC) based on the developed trapdoor one-way function (OWF). The public key is obtained by choosing a non-trivial generator $\gamma \in \mathbb{F}_{p}^{*}$. The chosen value of the generator constructs a cyclotomic matrix of order $2l^2$. It is believed that cyclotomic matrices of order $2l^2$ is always non-singular if the value of $k >1$. Since there are efficient algorithms for the construction of cyclotomic matrices. Consequently, the key setup time in our proposed cryptosystem is much shorter than previously designed/recently structured cryptosystems. 
\par
In our scheme, the secret key is given by choosing a different non-trivial generator, which is accomplished by discrete logarithm problem (DLP) over a finite field $\mathbb{F}_{p}^{*}$. A key-expansion algorithm is employed to expand the secret keys, which form a non-singular matrix of order $2l^2$. Here it is important to note that, if one can change the generators of $\mathbb{F}_{p}^{*}$, then entries of cyclotomic matrices get interchanged among themselves, however, the nature of the cyclotomic matrices remain the same. The decryption algorithm involves efficient algebraic operations of matrices. Hence the decryption in our proposed CAC is very efficient. In view of the perspective on the efficient encryption and decryption features, the polynomial time algorithm ensures that the proposed CAC makes it attractive in computationally restricted processors.
\par
The paper is organized as follows: Section \ref{sec1} presents the definition and notations, including some well-known properties of cyclotomic numbers of order $2l^{2}$. Section \ref{sec4} presents the construction of cyclotomic matrices of order $2l^{2}$. Section \ref{sec5} contains encryption and decryption algorithms of CAC along with a numerical example. In addition, the computational complexity of the proposed CAC is discussed step-wise in a mathematical language. Section \ref{sec7} presents the encryption and decryption can be efficiently perform with asymptotic complexity of $\mathcal{O}(e^{2.373})$. Finally, a brief conclusion is reflected in Section \ref{sec8}.
\section{Cyclotomic numbers}\label{sec1}
One of the central problems in the study of cyclotomic numbers is the determination
of all cyclotomic numbers of a specific order for a given finite field in terms of
solutions of certain Diophantine systems. Complete solutions
to the cyclotomy problem over a finite field $\mathbb{F}_{q}$ with characteristic $p$ have been
investigated by many authors for some specific orders. The problem of cyclotomy of order $2l^{2}$ concerns to formulate all $4l^{4}$ cyclotomic numbers of order $2l^{2}$. The section contains the generalized definition of cyclotomic numbers of order $e$, useful notations followed by properties of cyclotomic numbers of order $2l^{2}$. These properties play a major role in determining which cyclotomic numbers of order $2l^{2}$ are sufficient for the determination of all $4l^{4}$ cyclotomic numbers of order $2l^{2}$. The section also examines the cyclotomic matrices of order $2l^{2}$.
\subsection{Definition and notations} 
Let $e\geq 2$ be an integer, and $p\equiv 1 \pmod e$ an odd prime. One writes $p=ek+1$ for some positive integer $k$. Let $\mathbb{F}_{p}$ be the finite field of $p$ elements and let $\gamma$ be a generator of the cyclic group $\mathbb{F}_{p}^{*}$. For $0\leq a, b \leq e-1$, the cyclotomic number $(a,b)_{e}$ of order $e$ is defined as the number of solutions $(s,t)$ of the following:
\begin{equation} \label{1}
\gamma^{es+a}+\gamma^{et+b}+1\equiv 0 \pmod p; \ \ \ \ \ \ 0\leq s,t \leq k-1.
\end{equation} 
\subsection{Properties of cyclotomic numbers of order $2l^{2}$}\label{sec11}
 Let $p\equiv 1\pmod {2l^2}$ be a prime for an odd prime $l$ and we write $p=2l^2k+1$ for some positive integer $k$. It is clear that $(a,b)_{2l^2}=(a' ,b')_{2l^2}$ whenever $a\equiv a'\pmod{2l^2}$ and $b\equiv b'\pmod{2l^2}$ as well as $(a,b)_{2l^2}= (2l^2-a,b-a)_{2l^{2}}$. These imply the following:
\begin{equation}\label{2.1}
(a,b)_{2l^{2}}=\begin{cases}
 (b,a)_{2l^{2}}\hspace*{1.761cm} \text{ if }  k \text{ is  even},\\
(b+l^{2},a+l^{2})_{2l^{2}}\hspace*{0.461cm} \text{ if } k \text{ is odd}.
\end{cases} 
\end{equation}
Applying these facts, one can check that 
\begin{equation} \label{2.2}
\sum_{a=0}^{2l^{2}-1}\sum_{b=0}^{2l^{2}-1}(a,b)_{2l^{2}}=q-2
\end{equation}
and
\begin{equation} \label{2.3}
\sum_{b=0}^{2l^{2}-1}(a,b)_{2l^{2}}=k-n_{a},
\end{equation}
where $n_{a}$ is given by 
\begin{equation*}
n_{a}=\begin{cases}
1 \ \ \  \text{ if } a=0, 2\mid k \text{ or if } a=l^2, 2\nmid k;\\
0 \ \ \  \text{ otherwise }.
\end{cases}
\end{equation*}

\section{Cyclotomic Matrices} \label{sec4}
This section presents the procedure to determine cyclotomic matrices of order $2l^{2}$ for prime $l$. We determine the equality relation of cyclotomic numbers and discuss how few of the cyclotomic numbers are enough for the construction of whole cyclotomic matrix. Further generators for a chosen value of $p$ will be determined followed by the generation of a cyclotomic matrix. At every step, we have included a numerical example for the convenience to understand the procedure easily.
\par
\textbf{Definition:-} Cyclotomic matrix of order $2l^{2}$, $l$ be a prime, is a square matrix of order $2l^{2}$, whose entries are pair of solutions $(a,b)_{2l^{2}}$; $0\leq a,b\leq 2l^{2}-1$, of the equation (\ref{1}).

\begin{table}[h!]
\centering
\caption{Cyclotomic matrix of order 8}
\label{tab:cm8ke}
\begin{tabular}{@{}lllllllll@{}}
\toprule
(a,b) & \multicolumn{8}{c}{b}\\ \midrule
a     & \multicolumn{1}{c}{0} & \multicolumn{1}{c}{1} & \multicolumn{1}{c}{2} & \multicolumn{1}{c}{3} & \multicolumn{1}{c}{4} & \multicolumn{1}{c}{5} & \multicolumn{1}{c}{6} & \multicolumn{1}{c}{7} \\ \midrule
0 &(0,0)&(0,1)&(0,2)&(0,3)&(0,4)&(0,5)&(0,6)&(0,7)       \\
1 &(1,0)&(1,1)&(1,2)&(1,3)&(1,4)&(1,5)&(1,6)&(1,7)       \\
2 &(2,0)&(2,1)&(2,2)&(2,3)&(2,4)&(2,5)&(2,6)&(2,7)       \\
3 &(3,0)&(3,1)&(3,2)&(3,3)&(3,4)&(3,5)&(3,6)&(3,7)       \\
4 &(4,0)&(4,1)&(4,2)&(4,3)&(4,4)&(4,5)&(4,6)&(4,7)       \\
5 &(5,0)&(5,1)&(5,2)&(5,3)&(5,4)&(5,5)&(5,6)&(5,7)       \\
6 &(6,0)&(6,1)&(6,2)&(6,3)&(6,4)&(6,5)&(6,6)&(6,7) \\
7 &(7,0)&(7,1)&(7,2)&(7,3)&(7,4)&(7,5)&(7,6)&(7,7)\\ \bottomrule
\end{tabular}%
\end{table}
\par
For instance Table~\ref{tab:cm8ke} depicts a typical cyclotomic matrix of order 8 (assuming $l=2$). Whose construction steps have been given in the next subsection.
\subsection{Construction of cyclotomic matrix}
Typically construction of a cyclotomic matrix has been subdivided into four subsequent steps. Below are those ordered steps for the construction of a cyclotomic matrix;
\begin{enumerate}
\item For given $l$, choose a prime $p$ such that $p$ satisfies $p=2l^{2}k+1$, $k \in \mathbb{Z}^{+}$. The initial entries of the cyclotomic matrix are the arrangement of pair of numbers $(a,b)_{2l^{2}}$ where $a$ and $b$ usually vary from $0$ to $2l^{2}-1$.

\item Determine the equality relation of pair of $(a,b)_{2l^{2}}$, which reduces the complexity of pair of solution $(a,b)_{2l^{2}}$ of equation (\ref{1}), that is discuss in next sub-section.

\item Determine the generators of chosen $p$ (i.e. generators of $\mathbb{F}_{p}^{*}$). Let $\gamma_{1}$, $\gamma_{2}$, $\gamma_{3}$, \dots, $\gamma_{n}$ be generators of $\mathbb{F}_{p}^{*}$.

\item Choose a generator (say $\gamma_{1}$) of $\mathbb{F}_{p}^{*}$ and put in equation (\ref{1}). This will give cyclotomic matrix of order $2l^{2}$ w.r.t. chosen generator $\gamma_{1}$.  
\end{enumerate}

\par
The first step initializes the entries of cyclotomic matrix of order {2$l^2$}. Value of $p$ will be determined for given $l$. Assuming $l=2$, an example of such initialization of matrix of order 8 has been shown in Table~\ref{tab:cm8ke}. 
\par
For the construction of cyclotomic matrix, it does not require to determine all the cyclotomic numbers of a cyclotomic matrix which is shown in Table~\ref{tab:cm8ke} \cite{Helal2}. By well-known properties of cyclotomic numbers of order $2l^{2}$, cyclotomic numbers are divided into various classes, therefore there are a pair of the relation between the entries of initial table (Table~\ref{tab:cm8ke}) of a cyclotomic matrix. Thus to avoid calculating the same solutions in multiple times, we determine the equality relation of cyclotomic numbers (i.e. equality of solutions of $(a,b)_{2l^{2}}$). In the next subsection, we will discuss which cyclotomic numbers are enough for the construction of the cyclotomic matrix. Thus it helps us to the faster computation of cyclotomic matrix.
\subsection{Determination of equality relation of cyclotomic numbers}
This subsection presents the procedure to determine the equality relation of cyclotomic numbers (i.e. the relation of pair of $(a,b)_{2l^{2}}$), which reduces the complexity of solutions of pair of $(a,b)_{2l^{2}}$ (see also \cite{Helal2}). For the determination of cyclotomic matrices, it is not necessary to obtain all $4l^{4}$ cyclotomic numbers of order $2l^{2}$. 
The minimum number of cyclotomic numbers required to determine all the cyclotomic numbers (i.e. required for construction of cyclotomic matrix) depends on the value of positive integer $k$ on expressing prime $p=2l^{2}k+1$.
By (\ref{2.1}), if $k$ is even, then 
\begin{align} \label{equ8}
\nonumber (a,b)_{2l^{2}}& = 
(b,a)_{2l^{2}}=
(a-b,-b)_{2l^{2}}= 
(b-a,-a)_{2l^{2}}\\ & = 
(-a,b-a)_{2l^{2}}=
(-b,a-b)_{2l^{2}}
\end{align}
otherwise
\begin{align} \label{equ9}
\nonumber (a,b)_{2l^{2}}&= (b+l^{2},a+l^{2})_{2l^{2}}= (l^{2}+a-b,-b)_{2l^{2}}\\ & \nonumber = (l^{2}+b-a,l^{2}-a)_{2l^{2}} =(-a,b-a)_{2l^{2}}\\ & = (l^{2}-b,a-b)_{2l^{2}}.
\end{align}
Thus by (\ref{equ8}) and (\ref{equ9}), cyclotomic numbers $(a,b)_{2l^{2}}$ of order $2l^{2}$ can be divided into various classes. 

\begin{itemize}
\item $2|k$ and $l\neq 3$: In this case, (\ref{equ8}) gives classes of singleton, three and six elements. $(0,0)_{2l^{2}}$ form singleton class, $(-a,0)_{2l^{2}}$, $(a,a)_{2l^{2}}$, $(0,-a)_{2l^{2}}$ form classes of three elements where $1\leq a\leq {2l^{2}}-1 \pmod {{2l^{2}}}$ and rest $4l^{4}-3\times 2l^{2}+2$ of the cyclotomic numbers form classes of six elements.
\item $2|k$ and $l=3$: In this case, (\ref{equ8}) divide cyclotomic numbers $(a,b)_{18}$ of order $18$  into classes of singleton, second, three and six elements. $(0,0)_{18}$ form singleton class, $(-a,0)_{18}$, $(a,a)_{18}$, $(0,-a)_{18}$ form classes of three elements, where $1\leq a\leq 17 \pmod {18}$, $(6,12)_{18}=(12,6)_{18}$ which is grouped into classes of two elements and rest $4l^{4}-3\times 2l^{2}$ of the cyclotomic numbers form classes of six elements.
\item $2\nmid k$ and $l\neq 3$: Using (\ref{equ9}), once again we get  classes of singleton, three and six elements. $(0,l^{2})_{2l^{2}}$ forms singleton class, $(0,a)_{2l^{2}}$, $(a+l^{2},l^{2})_{2l^{2}}$, $(l^{2}-a,-a)_{2l^{2}}$ form classes of three elements, where $0\leq a \neq l^{2} \leq {2l^{2}}-1 \pmod {{2l^{2}}}$ and rest $4l^{4}-3\times 2l^{2}+2$ of the cyclotomic numbers form classes of six elements.
\item $2\nmid k$ and $l=3$: In this situation, (\ref{equ9}) partitions cyclotomic numbers $(a,b)_{18}$ of order $18$  into classes of singleton, two, three and six elements. Here $(0,9)_{18}$ form singleton class, $(0,a)_{18}$, $(a+9,9)_{18}$, $(9-a,-a)_{18}$ form classes of three elements, where $0\leq a\neq 9\leq 17 \pmod {18}$, $(6,3)_{18}=(12,15)_{18}$ which is grouped into classes of two elements and rest $4l^{4}-3\times 2l^{2}$ of the cyclotomic numbers form classes of six elements.
\end{itemize}
\begin{algorithm} [H]
\caption{Equality relation of cyclotomic numbers}
\label{algo1}
\begin{algorithmic}[1]
\vspace{.5cm}
\STATE START
\STATE Declare integer variable $e,l,p,k,flag$.
\STATE INPUT $l$, an odd prime and $e=2l^{2}$
\STATE Declare an array of size $e\times e$, where each element of array is $2$ tuple structure (i.e. ordered pair of $(a,b)_{2l^{2}}$, where $a$ and $b$ are integers).
\STATE INPUT $p$, prime number greater than 2
\IF{$(p-1)\%e==0$}
\STATE $k=(p-1)/e$
\IF{ $k$ even}
\STATE \textbf{Update table (E)}
\ELSE
\STATE \textbf{Update table (O)}
\ENDIF
\ENDIF
\end{algorithmic}
\end{algorithm}
Here \textbf{Update table (E)} means each entry $(a,b)_{2l^{2}}$ of the table will be updated by applying the relations $(a,b)_{2l^{2}}=(b,a)_{2l^{2}}=(a-b,-b)_{2l^{2}}=(b-a,-a)_{2l^{2}}=(-a,b-a)_{2l^{2}}=(-b,a-b)_{2l^{2}}$, and \textbf{Update table (O)} means each entry $(a,b)_{2l^{2}}$ of the table will be updated by applying the relations $(a,b)_{2l^{2}}=(b+l^{2},a+l^{2})_{2l^{2}}=(l^{2}+a-b,-b)_{2l^{2}}=(l^{2}+b-a,l^{2}-a_{2l^{2}})=(-a,b-a)_{2l^{2}}=(l^{2}-b,a-b)_{2l^{2}}$. 
\par
Further, if entries of the updated table are non-negative, then each entry should be replace by $\pmod {2l^{2}}$, otherwise add $\ 2l^{2}$. 
It is clear from above exploration, cyclotomic numbers of order $2l^{2}$ are divided into different classes depending on the values of $k$ and $l$. For $l=2$ and let $k$ be even, then $(0,0)_{8}$ give unique solution, cyclotomic numbers of the form $(-a,0)_{8}$, $(a,a)_{8}$, $(0,-a)_{8}$ where $1\leq a\leq 7 \pmod 8$ gives the same solutions and rest of cyclotomic numbers (i.e. $42$) which forms classes of six elements has maximum $7$ distinct numbers of solutions. Therefore the initial table (i.e. Table~\ref{tab:cm8ke}) of cyclotomic matrix reduces to Table~\ref{tab:cm8keven}. Similarly, for $l=2$ and let $k$ be odd, then $(0,4)_{8}$ give unique solution, cyclotomic numbers of the form $(0,a)_{8}$, $(a+4,4)_{8}$, $(4-a,-a)_{8}$ where $0\leq a \neq 4 \leq 7 \pmod {8}$ gives the same solutions and rest of cyclotomic numbers (i.e. $42$) which forms classes of six elements has maximum $7$ distinct numbers of solutions. Therefore the initial table (i.e. Table~\ref{tab:cm8ke}) of cyclotomic matrix reduces to Table~\ref{tab:cm8kodd}. One can observe that $64$ pairs of two parameter numbers $(a,b)_{8}$ reduced to $15$ distinct pairs (see Table~\ref{tab:cm8keven} and Table~\ref{tab:cm8kodd}). 
\begin{table}[h!]
\centering
\caption{Cyclotomic matrix of order 8 for even k}
\label{tab:cm8keven}
\begin{tabular}{@{}lllllllll@{}}
\toprule
(a,b) & \multicolumn{8}{c}{b}\\ \midrule
a     & \multicolumn{1}{c}{0} & \multicolumn{1}{c}{1} & \multicolumn{1}{c}{2} & \multicolumn{1}{c}{3} & \multicolumn{1}{c}{4} & \multicolumn{1}{c}{5} & \multicolumn{1}{c}{6} & \multicolumn{1}{c}{7} \\ \midrule
0 &(0,0)&(0,1)&(0,2)&(0,3)&(0,4)&(0,5)&(0,6)&(0,7)       \\
1 &(0,1)&(0,7)&(1,2)&(1,3)&(1,4)&(1,5)&(1,6)&(1,2)       \\
2 &(0,2)&(1,2)&(0,6)&(1,6)&(2,4)&(2,5)&(2,4)&(1,3)       \\
3 &(0,3)&(1,3)&(1,6)&(0,5)&(1,5)&(2,5)&(2,5)&(1,4)       \\
4 &(0,4)&(1,4)&(2,4)&(1,5)&(0,4)&(1,4)&(2,4)&(1,5)       \\
5 &(0,5)&(1,5)&(2,5)&(2,5)&(1,4)&(0,3)&(1,3)&(1,6)       \\
6 &(0,6)&(1,6)&(2,4)&(2,5)&(2,4)&(1,3)&(0,2)&(1,2) \\
7 &(0,7)&(1,2)&(1,3)&(1,4)&(1,5)&(1,6)&(1,2)&(0,1)\\ \bottomrule
\end{tabular}%
\end{table}
\begin{table}[h!]
\centering
\caption{Cyclotomic matrix of order 8 for odd k}
\label{tab:cm8kodd}
\begin{tabular}{@{}lllllllll@{}}
\toprule
(a,b) & \multicolumn{8}{c}{b}\\ \midrule
a     & \multicolumn{1}{c}{0} & \multicolumn{1}{c}{1} & \multicolumn{1}{c}{2} & \multicolumn{1}{c}{3} & \multicolumn{1}{c}{4} & \multicolumn{1}{c}{5} & \multicolumn{1}{c}{6} & \multicolumn{1}{c}{7} \\ \midrule
0 &(0,0)&(0,1)&(0,2)&(0,3)&(0,4)&(0,5)&(0,6)&(0,7)       \\
1 &(1,0)&(1,1)&(1,2)&(1,3)&(0,5)&(0,3)&(1,3)&(1,7)       \\
2 &(2,0)&(2,1)&(2,0)&(1,7)&(0,6)&(1,3)&(0,2)&(1,2)       \\
3 &(1,1)&(2,1)&(2,1)&(1,0)&(0,7)&(1,7)&(1,2)&(0,1)       \\
4 &(0,0)&(1,0)&(2,0)&(1,1)&(0,0)&(1,0)&(2,0)&(1,1)       \\
5 &(1,0)&(0,7)&(1,7)&(1,2)&(0,1)&(1,1)&(2,1)&(2,1)       \\
6 &(2,0)&(1,7)&(0,6)&(1,3)&(0,2)&(1,2)&(2,0)&(2,1) \\
7 &(1,1)&(1,2)&(1,3)&(0,5)&(0,3)&(1,3)&(1,7)&(1,0)\\ \bottomrule
\end{tabular}%
\end{table}

\par
\paragraph{\textbf{Remark 3.0}}By Algorithm \ref{algo1}, to compute $2l^{2}$ cyclotomic numbers, it is enough to compute $2l^{2}+\Big \lceil (2l^{2}-1)(2l^{2}-2)/6 \Big\rceil $, if $(2l^{2}-1)(2l^{2}-2)|6$, otherwise $2l^{2}+\Big \lceil (2l^{2}-1)(2l^{2}-2)/6 \Big\rceil +1$. Further, when $l$ is the least odd prime i.e. $l=3$, then $(2l^{2}-1)(2l^{2}-2)\nmid 6$. Therefore $l=3$, it is enough to calculate $64$ distinct cyclotomic numbers of order $2l^2$ and for $l\neq 3$, it is sufficient to calculate $2l^2+{(2l^2-1)(2l^2-2)}/6$ distinct cyclotomic numbers of order $2l^2$. 
\subsection{Determination of generators of $\mathbb{F}_{p}^{*}$}
To determine the solutions of (\ref{1}), we need the generator of the cyclic group $\mathbb{F}_{p}^{*}$. Let us choose finite field of order $p$ that satisfy $p=2l^{2}k+1; k\in \mathbb{Z^{+}}$. Let $\gamma_{1}$, $\gamma_{2}$, $\gamma_{3}$, \dots, $\gamma_{n}$ be generators of $\mathbb{F}_{p}^{*}$.
We  consider finite field of order $17$ (i.e. $\mathbb{F}_{17}$), since the chosen value of $p=17$ with respect to the value of $l$ take previously. Now to determine the generators of cyclic group $\mathbb{F}_{17}^{*}$. The detail procedure to obtain the generator of $\mathbb{F}_{17}^{*}$ has been depicted in Algorithm \ref{algo2}. If $G_{17}$ is a set that contain all the generator of $\mathbb{F}_{17}^{*}$, we could get elements of $G_{17}$ as $\{3$, $5$, $6$, $7$, $10$, $11$, $12$, $14\}$. 
\begin{algorithm}[h!]
\caption{Determination of generators of $\mathbb{F}_{p}^{*}$}
\label{algo2}
\begin{algorithmic}[1]
\vspace{.5cm}
\STATE Declare integer variable $p$, count
\STATE Declare integer array $arr\mathbb{F}_{p}[p],\ arr\mathbb{F}_{p}flag[p]$
\FOR {$i=1$ to $p-1$}
\STATE $arr\mathbb{F}_{p}[i]=i$, $arr\mathbb{F}_{p}flag[i]=0$ 
\ENDFOR
\STATE Declare integer array $arr\mathbb{G}_{p}[max]$
\STATE Declare integer variable $flag=0,\ r, \ \gamma$
\FOR {$i=1$ to $p-1$}
\STATE count=0
\FOR {$f=1$ to $p-1$}
\STATE $arr\mathbb{F}_{p}flag[f]=0$
\ENDFOR
\STATE $\gamma=arr\mathbb{F}_{p}[i]$
\FOR {$a=1$ to $p-1$}
\STATE $r=power(\gamma,a) \pmod p$
\FOR {$j=1$ to $p-1$}
\IF {$r$ is equal to $arr\mathbb{F}_{p}[j]$}
\STATE $arr\mathbb{F}_{p}flag[j]=1$
\ENDIF
\ENDFOR
\ENDFOR
\FOR {$k=1$ to $p-1$}
\IF {$arr\mathbb{F}_{p}flag[k]$ is equal to $1$}
\STATE count++ 
\ENDIF
\ENDFOR
\IF {count is equal to $p-1$}
\STATE $\gamma$ is generator
\ENDIF
\ENDFOR 
\end{algorithmic}
\end{algorithm}
\subsection{Generation of cyclotomic matrices}
This subsection, present an algorithm for the generation of cyclotomic matrices of order $2l^{2}$. Note that entries of cyclotomic matrices are solutions of (\ref{1}). Thus we need the generator of the cyclic group $\mathbb{F}_{p}^{*}$, which is discussed in the previous subsection.
On substituting the generators of $\mathbb{F}_{p}^{*}$ in Algorithm \ref{algo3}, we obtain the cyclotomic matrices of order $2l^{2}$ corresponding to different generators of $\mathbb{F}_{p}^{*}$. The chosen value of $p=17$ implies $k=2$ w.r.t. assume value of $l=2$. Therefore the cyclotomic matrix will be obtain from Table~\ref{tab:cm8keven}. Let us choose a generator (say $\gamma_{1}=3$) from set $G_{17}$. On substituting $\gamma_{1}=3$ in Algorithm \ref{algo3}, it will generate cyclotomic matrix of order $8$ over $\mathbb{F}_{17}$ w.r.t. chosen generator $\gamma_{1}=3$. Matrix $B_{0}$ show the corresponding cyclotomic matrix of order $8$ w.r.t. chosen generator $3\in \mathbb{F}_{17}^{*}$.
\begin{center}
	\textbf{B$_{0}$}
	{\small$
	=\left[
	\begin{array}{rrrrrrrr}
    0 & 0 & 0 & 0 & 0 & 0 & 1 & 0
	\\
	0 & 0 & 0 & 0 & 1 & 0 & 1 & 0 
	\\
	0 & 0 & 1 & 1 & 0 & 0 & 0 & 0 
	\\

	0 & 0 & 1 & 0 & 0 & 0 & 0 & 1 \\
	
	0 & 1 & 0 & 0 & 0 & 1 & 0 & 0 \\
	
	0 & 0 & 0 & 0 & 1 & 0 & 0 & 1 \\
	
	1 & 1 & 0 & 0 & 0 & 0 & 0 & 0 \\

    0 & 0 & 0 & 1 & 0 & 1 & 0 & 0 \\
	\end{array}\right] 
	$ }
\end{center}
\begin{algorithm}[h!]
\caption{Generation of cyclotomic matrix}
\label{algo3}
\begin{algorithmic}[1]
\vspace{.5cm}
\STATE INPUT: The value of $p, l,\gamma$
\STATE Declare an array $arr[ROW][COL]$ (where elements are two tuple structure)
\STATE Declare integer variable $p,$ $l,$ $k,$ $\gamma,$ $x,$ $y,$ $A,$ $s,$ $t,$ $a,$ $b,$ $count = 0,$ $p_{1},$ $p_{2}$
\FOR {a equal to $0$ to number of rows}
\FOR {b equal to $0$ to number of columns}
\FOR {$x$ is equal to $0$ to $k$}
\FOR {$y$ is equal to $0$ to $k$}
\STATE $p_{1}=2l^{2}*s+arr[a][b].l$
\STATE $p_{2}=2l^{2}*t+arr[a][b].m$
\STATE $A=power(\gamma,p_{1})+power(\gamma,p_{2})+1$
\IF {$A \pmod p$ is equal to $0$}
\STATE $count ++$
\ENDIF
\ENDFOR
\ENDFOR
\STATE $arr[a][b].n=count$
\STATE $count=0$
\ENDFOR
\ENDFOR  
\end{algorithmic}
\end{algorithm}  

\paragraph{\textbf{Remark 3.1}}
It is noted  that if we change the generator of $\mathbb{F}_{p}^{*}$, then entries of cyclotomic matrices get interchanged among themselves but their nature remains the same.
\paragraph{\textbf{Remark 3.2}}
It is obvious that (by (\ref{2.3})) cyclotomic matrices of order $2l^{2}$ is always singular if the value of $k=1$. 
\section{The public-key cryptosystem}
\label{sec5}
In this section, we present the approach for designing a public key cryptosystem using cyclotomic matrices discussed in section \ref{sec4}. The scheme employ matrices of order $2l^{2}$, whose entries are cyclotomic numbers of order $2l^{2}$. The public key is a non-trivial generator, say $\gamma^{\prime}$ of a set of generator in $\mathbb{F}_{p}^{*}$ along with $p$ and $l$. The set of generator is obtain by Algorithm \ref{algo2}. The chosen public keys generate a cyclotomic matrix as of required order (i.e. order of $2l^{2}$) make use of Algorithm \ref{algo3}.  Here, we define a trapdoor one-way function $\varPsi: \mathbb{F}_{p}^{*} \longrightarrow \mathbb{F}_{p}^{*}$ as $\varPsi(r_{0})=log_{\gamma^{\prime}}(\gamma^{\prime\prime})$; $r_{0} \in \mathbb{N}$, $\gamma^{\prime}, \gamma^{\prime\prime}$ are non-trivial generators of $\mathbb{F}_{p}^{*}$. Thus, the secret key are the values of $p$, $l$, $\gamma^{\prime\prime}$ \&  $r_{0}$. To encrypt a message, define composition of matrix as $M_{2l^{2}}(A*B)\longrightarrow M_{2l^{2}}(C)$, where $A$ is a message block matrix, $B$ is a cyclotomic matrix w.r.t. $\gamma^{\prime} \in \mathbb{F}_{p}^{*}$ and $C$ is the ciphertext matrix. Other way one can define $M_{2l^{2}}(B*A)\longrightarrow M_{2l^{2}}(C)$. Therefore, the length of the ciphertext in CAC is equal to $2l^{2}$. 
\par
To decrypt a message, an algorithm is required to expand the secret keys provided by the secret values. Therefore, the Algorithm \ref{algo4} is utilized for this purpose.
\begin{algorithm}[H]
\caption{Secrete key expansion}
\label{algo4}
\begin{algorithmic}[1]
\vspace{.5cm}
\STATE SECRET INPUT: The values of $p$, $l$, $r_{0}$ and $\gamma^{\prime\prime}$
\STATE Algorithm \ref{algo1}
\STATE Algorithm \ref{algo3} 
\end{algorithmic}
\end{algorithm}
The main purpose, to utilize the above algorithm is to construct a non-singular cyclotomic matrix of order $2l^{2}$ w.r.t. non-trivial generator $\gamma^{\prime\prime}$ ($\gamma^{\prime\prime}\neq\gamma^{\prime}$) in $\mathbb{F}_{p}^{*}$. Now to decrypt the message, we define inverse composition relation of matrices, which is $M_{2l^{2}}(C*Z)\longrightarrow M_{2l^{2}}(A)$, where matrix $Z$ is obtain by some efficient algebraic computation of matrix. Other way one can define $M_{2l^{2}}(Z*C)\longrightarrow M_{2l^{2}}(A)$ respectively.  
\subsection{Determination of matrix $Z$}
The following steps have been taken for the determination of matrix $Z$.
\begin{enumerate}
    \item Determine the equality of cyclotomic matrix of order $2l^2$ corresponding to the secret values of $p$ \& $l$, which is perform by Algorithm \ref{algo1}.
    \item Each entry of equality of cyclotomic matrix is multiplied by $r_{0}$.
    \item Compute the inverse of equality of cyclotomic matrix generated in step 2.
    \item Finally, on substitution the values of the generated cyclotomic matrix corresponding to $\gamma^{\prime\prime}$ to an inverse matrix in step 3.
\end{enumerate}
\par
The following two algorithms (i.e. Algorithm \ref{algo5} \& \ref{algo6}) are utilized to encrypt and decrypt a message in the proposed CAC, respectively.
\begin{algorithm}[H]
\caption{Encryption}
\label{algo5}
\begin{algorithmic}[1]
\vspace{.5cm}
\STATE Transfer the plain text (message) into its numerical value and store in matrix of order $2l^{2}$
\STATE PUBLIC INPUT: The values of $p,l$ and $\gamma^{\prime}$
\STATE Execute Algorithm
\ref{algo3}
\STATE Check: Generated cyclotomic matrix in step 3 is non-singular
\STATE Cipher matrix: Multiply cyclotomic matrix and the matrix generated in step 1
\STATE Ciphertext: The corresponding text values of matrix generated in step 5
\end{algorithmic}
\end{algorithm}
\begin{algorithm}[H]
\caption{Decryption}
\label{algo6}
\begin{algorithmic}[1]
\vspace{.5cm}
\STATE Input: The cipher matrix/ciphertext
\STATE Execute Algorithm \ref{algo4}
\STATE Each entries of equality of cyclotomic matrix (i.e. output matrix of Algorithm \ref{algo1}) is multiply by $r_{0}$. The entries of the generated matrix are pair of cyclotomic number
\STATE Compute the inverse of generated matrix in step $3$ and substitute the value of each pair of cyclotomic number from generated matrix in step $2$ 
\STATE Now multiply the cipher text matrix to generated matrix in step $4$, we get back to the original plain text message.  
\end{algorithmic}
\end{algorithm}
\subsection{Computational complexity of the CAC}
In this section, we would validate the computational complexity of the proposed CAC. The computational complexity measures the amount of computational effort required, by the best as of now known techniques, to break a system \cite{Menezes1}. However, it is exceptionally hard to demonstrate the computational complexity of public-key cryptosystems \cite{Menezes1, Stinson1}. For instance, if the public modulus of RSA is factored into its prime components, at that point the RSA is broken. Be that as it may, it isn't demonstrated that breaking RSA is identical to factoring its modulus \cite{57X}. Here, we study the computational complexity of the CAC by providing arguments related to the inversion of the one-way function in CAC to a best known computational algorithm. The complexity of anonymous decryption could be understood as; if we assume that an attacker wants to recover the secret key by using all the information's available to them. Then they need to solve the discrete logarithm problem (DLP) to find the secret key followed by a number of steps described in Algorithm~\ref{algo6}. Since, the one-way function is define analogous to discrete logarithm problem (DLP). However, although most mathematicians and computer scientists believe that the DLP is unsolvable \cite{AAA}. The complexity of the DLP depends on the cyclic group. It is believed to be a hard problem for the multiplicative group of a finite field of large cardinality. Therefore even determining the very first step is nearly unsolvable. 
\par
If it is the case that somehow attacker manages to solve the DLP, then they have to determine equation (\ref{1}) and calculate all the solutions corresponding to different pairs $(a,b)_{2l^{2}}$. Further, it is required to determine the relation matrix based on equality relation among the solutions of equation (\ref{1}). Where entries of the relation matrix are the two-tuple structure of $(a,b)_{2l^{2}}$. Finally, entries of inverse of the relation matrix are required to replace through the implication of DLP.
\par
Here we could observe the computational complexity as it increases with the value of $p$ and $2l^{2}$. Therefore it is nearly impossible to determine the secret key for a large value of $p$ and $2l^{2}$; hence uphold the secure formulation claim of the proposed work.
\par
\subsection{An example of the CAC}
In this section, we provide an example for the proposed CAC. The example is designed according to guidelines described in section \ref{sec5}. The main purpose of this example is to show the reliability of our cryptosystem. It is important to note that this example is non-viable for the proposed CAC, since the values of the parameters are too small. 
\par
Let us consider $2l^{2}=8$ (i.e. $l=2$) and $p=17$. Suppose we want to send a message $X$ whose numerical value store in matrix $\textbf{A}$ of order $8$.
\begin{center}
	\textbf{A}
	{\small$
	=\left[
	\begin{array}{rrrrrrrr}
    2 & 3 & 5 & 9 & 8 & 0 & 2 & 1
	\\
	1 & 5 & 9 & 2 & 9 & 3 & 0 & 5 
	\\
	2 & 1 & 3 & 2 & 5 & 6 & 8 & 7 
	\\

	5 & 3 & 0 & 7 & 8 & 7 & 3 & 1 \\
	
	4 & 2 & 3 & 1 & 9 & 8 & 7 & 3 \\
	
	0 & 9 & 2 & 3 & 5 & 6 & 8 & 9 \\
	
	1 & 0 & 2 & 9 & 6 & 7 & 9 & 8 \\

    9 & 1 & 3 & 2 & 4 & 4 & 5 & 6 \\
	\end{array}\right] 
	 $}
\end{center}
We choose two distinct non-trivial generators of a set of generator in $\mathbb{F}_{17}^{*}$ (the set of generator is obtain by employing Algorithm \ref{algo2}), say $\gamma^{\prime}=11$ and $\gamma^{\prime \prime}=3$. Now, we evaluate the complex relation between these chosen generators, which can perform by DLP. One can write $3^{7}=11$ $\pmod {17}$. Consider that $r_{0}=7$. The public key is the public values $l=2, \ p=17$ \& $\gamma^{\prime}=11$ and the private key is the secret values $l=2, \ p=17$, $r_{0}=7$ \& $\gamma^{\prime\prime}=3$. The public values generated cyclotomic matrix of order $8$ as required, which is 

\begin{center}
	\textbf{B$_{3}$}
	{\small$
	=\left[
	\begin{array}{rrrrrrrr}
    0 & 0 & 1 & 0 & 0 & 0 & 0 & 0 \\
	0 & 0 & 0 & 1 & 0 & 1 & 0 & 0 \\
	1 & 0 & 0 & 0 & 0 & 0 & 0 & 1 \\
        
	0 & 1 & 0 & 0 & 1 & 0 & 0 & 0 \\
	
	0 & 0 & 0 & 1 & 0 & 0 & 0 & 1 \\
	
	0 & 1 & 0 & 0 & 0 & 0 & 1 & 0 \\
	
	0 & 0 & 0 & 0 & 0 & 1 & 1 & 0 \\

    0 & 0 & 1 & 0 & 1 & 0 & 0 & 0 \\
	\end{array}\right] 
	$ }
\end{center} 
Determinant of \textbf{B}$_{3}$ is equal to $1$, implies non-singular. Now we encrypt the message $\textbf{A}$ by multiplying matrix \textbf{B}$_{3}$ and $\textbf{A}$, which is as follows:
\begin{center}
	\textbf{C=B$_{3}\times$ A}
	{\small$
	=\left[
	\begin{array}{rrrrrrrr}
    2 & 1 & 3 & 2 & 5 & 6 & 8 & 7
	\\
	5 & 12 & 2 & 10 & 13 & 13 & 11 & 10 
	\\
	11 & 4 & 8 & 11 & 12 & 4 & 7 & 7 
	\\

	5 & 7 & 12 & 3 & 18 & 11 & 7 & 8 \\
	
	14 & 4 & 3 & 9 & 12 & 11 & 8 & 7 \\
	
	2 & 5 & 11 & 11 & 15 & 10 & 9 & 13 \\
	
	1 & 9 & 4 & 12 & 11 & 13 & 17 & 17 \\

    6 & 3 & 6 & 3 & 14 & 14 & 15 & 10 \\
	\end{array}\right] 
	$ }
\end{center}
The matrix \textbf{C} is a ciphertext matrix. To transmit the message, entries of the matrix converted into text. To decrypt the message, first, we expand the secret keys which are performed by Algorithm \ref{algo4}. It generates a non-singular cyclotomic matrix of order $8$, which is shown by matrix \textbf{B}$_{0}$. Now each entry of equality of cyclotomic matrix (i.e. output matrix of Algorithm \ref{algo1}) is multiplied by $r_{0}=7$. We get matrix $\textbf{D}$ whose entries are pair of cyclotomic numbers. 
\begin{center}
	\textbf{D}
	{\small$
	=\left[
	\begin{array}{rrrrrrrr} (0,0)&(0,7)&(0,6)&(0,5)&(0,4)&(0,3)&(0,2)&(0,1)
\\ (0,7)&(0,1)&(1,2)&(1,6)&(1,5)&(1,4)&(1,3)&(1,2)
\\  (0,6)&(1,2)&(0,2)&(1,3)&(2,4)&(2,5)&(2,4)&(1,6)
\\  (0,5)&(1,6)&(1,3)&(0,3)&(1,4)&(2,5)&(2,5)&(1,5)
\\(0,4)&(1,5)&(2,4)&(1,4)&(0,4)&(1,5)&(2,4)&(1,4)
\\ (0,3)&(1,4)&(2,5)&(2,5)&(1,5)&(0,5)&(1,6)&(1,3)
\\ (0,2)&(1,3)&(2,4)&(2,5)&(2,4)&(1,6)&(0,6)&(1,2)
\\ (0,1)&(1,2)&(1,6)&(1,5)&(1,4)&(1,3)&(1,2)&(0,7)
\\
	\end{array}\right] 
	$ }
\end{center} 
Now compute the inverse of $\textbf{D}$ and substitute the value from \textbf{B}$_{0}$ to each pair of cyclotomic numbers. The  matrix becomes
\begin{center}
\textbf{D$^{*}$}
{\small$
=\left[
\begin{array}{rrrrrrrr}
    -1 & 1 & 1 & -1 & -1 & 1 & -1 & 1
	\\
	1 & 0 & 0 & 1 & 0 & 0 & 0 & -1 
	\\
	1 & 0 & 0 & 0 & 0 & 0 & 0 & 0 
	\\

	-1 & 1 & 0 & -1 & 0 & 1 & -1 & 1 \\
	
	-1 & 0 & 0 & 0 & 0 & 0 & 0 & 1 \\
	
	1 & 0 & 0 & 1 & 0 & -1 & 1 & -1 \\
	
	-1 & 0 & 0 & -1 & 0 & 1 & 0 & 1 \\

    1 & -1 & 0 & 1 & 1 & -1 & 1 & -1 \\
	\end{array}\right] 
	 $}
\end{center} 

Finally, we obtain \textbf{D$^{*}\times$ C\ =\ A}.

\section{The complexity of CAC}\label{sec7}
Time and space are usually prominent factors to establish the effectiveness of security solutions. In the before seen sections, we have established the computation difficulty to break the presented work. Further, we would demonstrate the complexity of the solution in terms of worst-case running time. 

\par
The time complexity of Algorithm \ref{algo1} in worst case is $\mathcal{O}(e^2)$. Since creation of matrix of order $e$ and Update\_Table() individually will take $\mathcal{O}(e^2)$. In algorithm 2, \textbf{for} loop in line number 9, 15, and 17 contributes $\mathcal{O}(e^3)$ in worst case. Since,
\begin{equation*}
    e = \frac{p-1}{k}
\end{equation*}
\begin{equation*}
    e^3 = \bigg(\frac{p-1}{k}\bigg)^3 \equiv \bigg(\frac{p^3}{k^3}\bigg)
\end{equation*}
Since $k$ is a positive integer, therefore when $k$ attains its minimum value i.e. 1,
\begin{equation*}
\frac{p^3}{k^3} \equiv p^3 \equiv e^3   
\end{equation*}
For any higher value of $k$, there is guarantee that 
\begin{equation*}
\frac{p^3}{k^3} < e^3
\end{equation*}
Hence we can conclude that Algorithm \ref{algo2} can take $\mathcal{O}(e^3)$  in worst case. 
\par
Similarly in Algorithm \ref{algo3}, \textbf{for} loop in line number 4, 5, 6, 7 contributes $e.\ e.\ k.\ k$ or say  $\mathcal{O}(e^{2}k^{2})$ running time in worst case. Using similar analogy as in case of Algorithm \ref{algo2}, worst case complexity will be $\mathcal{O}(e^2)$.
\subsection{Encryption}
Encryption as expressed in Algorithm~\ref{algo5} constitutes of three logical divisions and the complexity of encryption would be the sum of the complexity of its part. The state divisions within are as follows;
\begin{enumerate}
    \item Generating cyclotomic matrix
    \item Checking the singularity of the cyclotomic matrix.
    \item Multiplication of generated cyclotomic matrix and matrix corresponds to plain text.
\end{enumerate}

Starting from the generation of the cyclotomic matrix, comprises the total complexity $\mathcal{O}(e^2)$ as stated earlier. Further, checking singularity involves the computation of determinants of the matrix of order $e$. In worst case computing determinant of a matrix of order $e$ by fast algorithm \cite{Aho} takes $\mathcal{O}(n^{2.373})$. Hence singularity of the cyclotomic matrix could be computed in $\mathcal{O}(e^{2.373})$ time.
Finally multiplication of cyclotomic matrix of order $e$ and matrix corresponds to plain text of order $e$ will take $\mathcal{O}(e^{2.3728639})$ time. Therefore, Complexity of Encryption would become;  $\mathcal{O}(e^2) + \mathcal{O}(e^{2.373}) + \mathcal{O}(e^{2.3728639}) \equiv  \mathcal{O}(e^{2.373})$. Finally a polynomial time complexity seems to be quite worthwhile.
\subsection{Decryption}
Decryption as expressed in Algorithm \ref{algo6} initially imply Algorithm \ref{algo4} which sums the complexity of Algorithm \ref{algo1} and \ref{algo3}, therefore takes $\mathcal{O}(e^2)$ + $\mathcal{O}(e^2) \equiv \mathcal{O}(e^2)$ time. Further, multiplication of cyclotomic matrix of order $e$ by a constant value $r_0$, therefore yield $\mathcal{O}(e^2)$ complexity. Inverse of a matrix of order $e$ can be computed by a fast algorithm \cite{Aho} in $\mathcal{O}(n^{2.373})$, therefore in our case inverse of generated matrix of order $e$ could be computed in $\mathcal{O}(e^{2.373})$ time. Finally multiplication of two matrix of order $e$ could be computed in $\mathcal{O}(e^{2.3728639})$ by best known algorithm\cite{Gall} till date. Therefore, Complexity of decryption would be $\mathcal{O}(e^2)$ + $\mathcal{O}(e^{2})$ + $\mathcal{O}(e^{2.373})$ + $\mathcal{O}(e^{2.3728639})$, which becomes $\mathcal{O}(e^{2.373})$.

\par
\section{Conclusion} \label{sec8}
In this paper, we have introduced a secured asymmetric key cryptography model applying the principle of cyclotomic numbers over a finite field. Procedure to generate cyclotomic matrix along with public \& private key have been presented where the relation between the public \& private key has acquired by discrete logarithm problem (DLP). Finally, a convincing argument to strengthen the claim has been presented followed by the method of encryption, decryption, and a numerical example. 

\end{document}